\documentclass[amsmath,amssymb,aps,prc,twocolumn,10pt]{revtex4}
\usepackage{graphicx}
\usepackage{dcolumn}
\usepackage{color}
\usepackage{bm}

\begin{document}

\preprint{APS}

\title{Light Nuclei embedded in a Nuclear Medium: Clustering and Mott Transitions}

\author{Qi Meng}
\author{Chang Xu}
 \email{Contact author: cxu@nju.edu.cn}
 \affiliation{School of Physics, Nanjing University, Nanjing 210093, China}
\date{\today}
\begin{abstract}
The clustering of nucleons is a fundamental phenomenon with broad implications for nuclear physics and astrophysics.
In this work, we employ a microscopic in-medium few-body approach to systematically investigate the formation and dissolution of light clusters (deuteron, $\rm{^3H}$, $\rm{^3He}$, $\alpha$-particle) embedded in nuclear medium. The medium-modified cluster structures under the Pauli blocking and the picture of Mott transitions in nuclear medium are discussed in detail.
We find that the weakly bound deuteron survives to higher densities as compared with the more compact $\alpha$-particle in symmetric nuclear matter, with its r.m.s. radius expanding markedly prior to its dissolution. Moreover, the Mott density of $\alpha$-particle is slightly lower in neutron-rich matter than in symmetric matter.
These results may provide useful constraints for the formation of light clusters at nuclear surface and the cluster yields in intermediate-energy heavy-ion collisions.

\end{abstract}

\maketitle
\section{Introduction}
Whereas the properties of light clusters in free space are relatively well understood, their behavior when embedded in the nuclear medium remains challenging to investigate both experimentally and theoretically \cite{tohsaki2017colloquium,Freer:2017gip,Ebran:2012ww,wei2024clustering}.
The presence of surrounding nucleons alters the correlations among the nucleons within clusters, primarily due to nucleon-nucleon interactions and Pauli blocking effects. The binding of the clusters is gradually reduced as the nucleon density of medium increases, leading eventually to their dissolution into uncorrelated nucleons at a critical density known as the Mott transition \cite{Ropke:1982ino,Ropke:1983lbc,Hagel:2011ws}.

The in-medium properties of light clusters are of relevance to several important issues including $\alpha$-cluster formation and decay in heavy nuclei \cite{Delion:1992zz,Buck:1992zz,Tohsaki:2001an,Royer:2002kga,Mohr:2006qm,Xu:2006fq,Chernykh:2007zz,Denisov:2009ng,Ropke:2014wsa,Xu:2017vyt,Ropke:2017qck,yang2021alpha,yang2023alpha} and light cluster production in intermediate-energy heavy-ion collisions (HIC) \cite{Kowalski:2006ju,Natowitz:2010ti,Typel:2009sy,Qin:2011qp,Pais:2019jst,Wang:2023gta}.
A precise understanding of these properties is essential for quantifying the $\alpha$-cluster preformation probability, which is a key factor in estimating $\alpha$-decay rates.
Microscopic studies have shown that the $\alpha$-cluster can appear only at the surface region of the core nucleus and the Mott transition from unbound four-nucleon shell-model state to quartet state occurs at about 1/5 normal density, $i.e.$, the Mott density \cite{Ropke:2014wsa,yang2021alpha,yang2023alpha}. The concept of the Mott transition is also critical in intermediate-energy HIC.
Recent kinetic-approach analyses indicate that refined Mott densities of light clusters are vital for explaining the observed enhancement of the $\alpha$-particle yield at low incident energies in HIC \cite{Wang:2023gta,FOPI:2010xrt,INDRA:2021fcr}.
Furthermore, clustering and Mott transition have significant implications for astrophysics of compact objects, in particular the properties of neutron star crusts and the astrophysical nucleosynthesis \cite{Sumiyoshi:2008qv,Raduta:2010ym,Aymard:2014hna,Oertel:2016bki}.

The formation and dissolution of light clusters in nuclear medium are rather complex to handle as one needs to treat not only the interactions between the nucleons in the cluster and in the medium but also the Pauli blocking effect arising from the surrounding nuclear medium. So far, only a few microscopic approaches have been devoted to the study of in-medium properties of light clusters. Using a quantum statistical approach, microscopic studies have examined the influence of the nuclear medium on the cluster abundances at finite temperature, revealing that the clusters are destructed once the density exceeds the Mott density \cite{Ropke:1982ino,Ropke:1983lbc}.
For the $\alpha$-particle embedded in homogeneous matter, the in-medium quantum four-body equation has been solved via variational calculations with a Gaussian ansatz \cite{Ropke:2014wsa}. In the context of finite nuclei, the quartetting wave function approach has been proposed to calculate microscopically the formation probability of an $\alpha$-cluster on the surface of the core and its decay rate  \cite{Xu:2006fq,Xu:2017vyt,yang2021alpha,yang2023alpha}.

In the present work, we refine a microscopic in-medium few-body approach to study the properties of light clusters with $A$=2-4 ($i.e.$, deuteron, $\rm{^3H}$, $\rm{^3He}$, and $\alpha$-particle) in nuclear matters. Firstly, the full quantum few-body equations are solved by expanding the trial wave function using a superposition of multi-Gaussian basis functions. The Gaussian range parameters are chosen to follow a geometric progression, a scheme proven highly effective in capturing both short-range correlations and long-range asymptotic behavior of few-body wave functions \cite{Hiyama:2003cu}. This method has previously been successfully applied to describe light clusters in free space \cite{Meng:2023age}.
Secondly, we perform more sophisticated calculations by considering different Fermi surfaces and Pauli blockings  for $\alpha$-particle in imbalanced nuclear matter. The Mott densities are systematically mapped for deuteron, $\rm{^3H}$, $\rm{^3He}$, and $\alpha$-particle. These microscopic results may allow us to provide a description of the Mott transitions for light clusters in nuclear medium in a unified framework.

The structure of this article is as follows: Sec.~\ref{framework} is the framework of the in-medium few-body approach with multi-Gaussian bases. The numerical results of intrinsic energy shifts and Mott densities are presented and discussed in Sec.~\ref{inmedium}. A brief summary is given in Sec.~\ref{summary}.

\section{Framework}
\label{framework}

\begin{figure}[ht]
\includegraphics[width=8.cm]{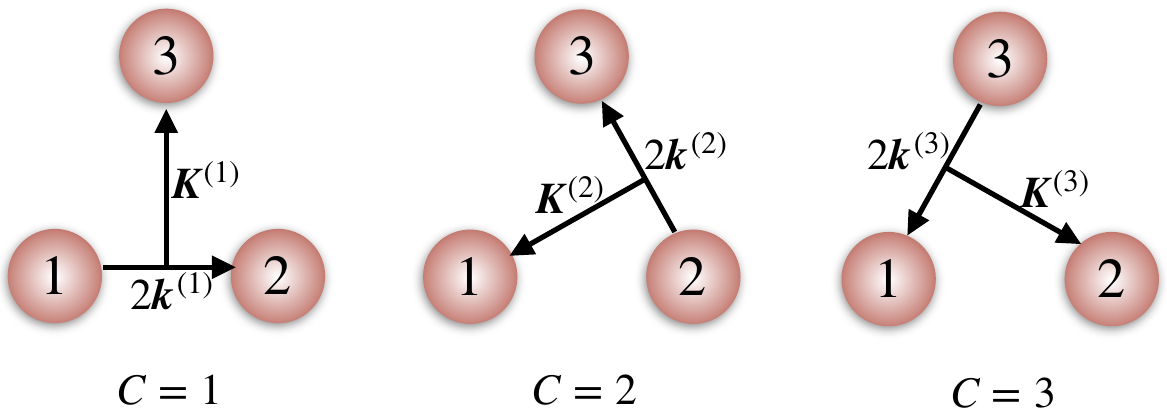}

\caption{\label{jacobiK3} Rearrangement channels of 3-nucleon cluster Jacobi momenta.}
\end{figure}

\begin{figure}[ht]
\includegraphics[width=8.3cm]{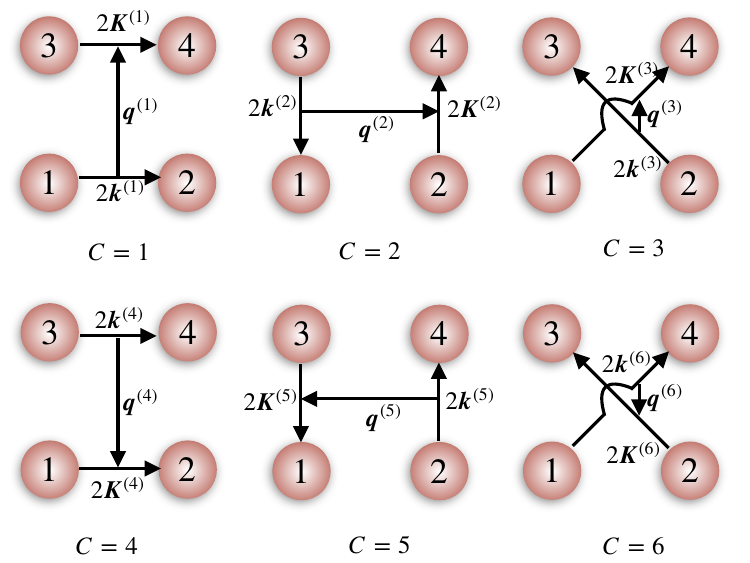}
\caption{\label{jacobiK4} Rearrangement channels of 4-nucleon cluster Jacobi momenta.}
\end{figure}

We consider an $A$-nucleon cluster embedded in homogeneous nuclear matter, which is characterized by the nucleon densities $n_b=n_n+n_p$ and the isospin asymmetry $\delta=(n_n-n_p)/n_b$, with $n_n$ and $n_p$ specifying the neutron and proton densities, respectively. Since the total momentum $\boldsymbol{P}$ of the cluster is conserved in homogeneous matter, the total wave function of the cluster with total angular momentum $J$ and its $z$-component $M$, and total isospin $T$ and its $z$-component $T_z$, can be decomposed into a center-of-mass (c.m.) part $\Psi_{JMTT_z}^{A,\rm{c.m.}}$ and an intrinsic part $\Psi_{JMTT_z}^{A,\rm{intr}}$.
In this work, we restrict our consideration to the case of zero temperature and zero total momentum $\boldsymbol{P}=\sum_{i=1}^A\boldsymbol{p}_i=0$, where $\boldsymbol{p}_{i}$ is the momentum of the $i$-th nucleon.
Then the energy of c.m. motion vanishes and we can write the in-medium Schr{\"o}dinger equation for the intrinsic motion in momentum space,
\begin{widetext}

\begin{equation}
\begin{aligned}
\label{schrodinger2}
\sum_{i=1}^A \frac{\boldsymbol{p}_i^2}{2m} &\Psi_{JMTT_z}^{A,\rm{intr}} (\boldsymbol{\kappa}_1^{(1)},\boldsymbol{\kappa}_2^{(1)},...,\boldsymbol{\kappa}_{A-1}^{(1)},...,\boldsymbol{\kappa}_{1}^{(C)},...,\boldsymbol{\kappa}_1^{(\binom{A}{2})},\boldsymbol{\kappa}_2^{(\binom{A}{2})},...,\boldsymbol{\kappa}_{A-1}^{(\binom{A}{2})}) \\
&+\sum_{i=1}^A V^{\rm{MF}}(\boldsymbol{p}_i) \Psi_{JMTT_z}^{A,\rm{intr}} (\boldsymbol{\kappa}_1^{(1)},\boldsymbol{\kappa}_2^{(1)},...,\boldsymbol{\kappa}_{A-1}^{(1)},...,\boldsymbol{\kappa}_{1}^{(C)},...,\boldsymbol{\kappa}_1^{(\binom{A}{2})},\boldsymbol{\kappa}_2^{(\binom{A}{2})},...,\boldsymbol{\kappa}_{A-1}^{(\binom{A}{2})})
\\& +\sum_{C=1}^{\binom{A}{2}} \int d \boldsymbol{\kappa}_1^{(C)\prime} V^{\rm{NN}}
(\boldsymbol{\kappa}_1^{(C)},\boldsymbol{\kappa}_1^{(C)\prime})
\Psi_{JMTT_z}^{A,\rm{intr}} (\boldsymbol{\kappa}_1^{(1)},\boldsymbol{\kappa}_2^{(1)},...,\boldsymbol{\kappa}_{A-1}^{(1)},...,\boldsymbol{\kappa}_{1}^{(C)\prime},...,\boldsymbol{\kappa}_1^{(\binom{A}{2})},\boldsymbol{\kappa}_2^{(\binom{A}{2})},...,\boldsymbol{\kappa}_{A-1}^{(\binom{A}{2})}) \\
&= E^{\rm{intr}} \Psi_{JMTT_z}^{A,\rm{intr}}
(\boldsymbol{\kappa}_1^{(1)},\boldsymbol{\kappa}_2^{(1)},...,\boldsymbol{\kappa}_{A-1}^{(1)},...,\boldsymbol{\kappa}_{1}^{(C)},...,\boldsymbol{\kappa}_1^{(\binom{A}{2})},\boldsymbol{\kappa}_2^{(\binom{A}{2})},...,\boldsymbol{\kappa}_{A-1}^{(\binom{A}{2})}),
\end{aligned}
\end{equation}
where $m$ is the mass of the nucleon.
$\boldsymbol{\kappa}_1^{(C)},\boldsymbol{\kappa}_2^{(C)},...,\boldsymbol{\kappa}_{A-1}^{(C)}$ are the relative momenta between the nucleons in the cluster, which can be defined in a set of rearrangement channels of Jacobi momenta $C=1$ to $\binom{A}{2}$. For the 2-nucleon cluster, the relative momentum is simply $\boldsymbol{\kappa}_1^{(1)}=\boldsymbol{k}=(\boldsymbol{p}_2-\boldsymbol{p}_1)/2$. For the 3- and 4-nucleon clusters, $(\boldsymbol{\kappa}_1^{(C)},\boldsymbol{\kappa}_2^{(C)}, \boldsymbol{\kappa}_3^{(C)})$ are denoted as $(\boldsymbol{k}^{(C)},\boldsymbol{K}^{(C)}, \boldsymbol{q}^{(C)})$ in Jacobi momenta, as shown in Fig.~\ref{jacobiK3} and Fig.~\ref{jacobiK4}, respectively.
For instance, the equation for the intrinsic motion of the 4-nucleon cluster is
\begin{equation}
\begin{aligned}
\label{schrodinger3}
\frac{\hbar^2}{2m}&\Big( 2(\boldsymbol{k}^{(1)})^2+2(\boldsymbol{K}^{(1)})^2+(\boldsymbol{q}^{(1)})^2 \Big) \Psi_{JMTT_z}^{A,\rm{intr}} (\boldsymbol{k}^{(1)},\boldsymbol{K}^{(1)},\boldsymbol{q}^{(1)},...,\boldsymbol{k}^{(C)},...,\boldsymbol{k}^{(\binom{A}{2})},\boldsymbol{K}^{(\binom{A}{2})},\boldsymbol{q}^{(\binom{A}{2})}) \\
&+\sum_{i=1}^A V^{\rm{MF}}(\boldsymbol{p}_i) \Psi_{JMTT_z}^{A,\rm{intr}} (\boldsymbol{k}^{(1)},\boldsymbol{K}^{(1)},\boldsymbol{q}^{(1)},...,\boldsymbol{k}^{(C)},...,\boldsymbol{k}^{(\binom{A}{2})},\boldsymbol{K}^{(\binom{A}{2})},\boldsymbol{q}^{(\binom{A}{2})})
\\& +\sum_{C=1}^{\binom{A}{2}} \int d \boldsymbol{k}^{(C)\prime} V^{\rm{NN}}
(\boldsymbol{k}^{(C)},\boldsymbol{k}^{(C)\prime}) \Psi_{JMTT_z}^{A,\rm{intr}} (\boldsymbol{k}^{(1)},\boldsymbol{K}^{(1)},\boldsymbol{q}^{(1)},...,\boldsymbol{k}^{(C)\prime},...,\boldsymbol{k}^{(\binom{A}{2})},\boldsymbol{K}^{(\binom{A}{2})},\boldsymbol{q}^{(\binom{A}{2})})  \\
&= E^{\rm{intr}} \Psi_{JMTT_z}^{A,\rm{intr}} (\boldsymbol{k}^{(1)},\boldsymbol{K}^{(1)},\boldsymbol{q}^{(1)},...,\boldsymbol{k}^{(C)},...,\boldsymbol{k}^{(\binom{A}{2})},\boldsymbol{K}^{(\binom{A}{2})},\boldsymbol{q}^{(\binom{A}{2})}).
\end{aligned}
\end{equation}

$E^{\rm{intr}}$ denotes the intrinsic energy of the cluster, which is shifted by the surrounding medium, including an external part and an internal part \cite{Ropke:2014wsa},
\begin{equation}
\begin{aligned}
\label{Eintrinsic}
E^{\rm{intr}}=U^{\rm{ext}}+U^{\rm{int}}.
\end{aligned}
\end{equation}
The external part, $U^{\rm{ext}}$, arises from interactions with the surrounding nucleons, which can be approximated by a mean-field potential $V^{\rm{MF}}(\boldsymbol{p}_i)$. This potential acts on both the nucleons in the cluster and the uncorrelated nucleons in the scattering states.
For simplicity, we neglect the momentum dependence of the mean-field potential, so that the energy shifts due to the mean field are the same for the bound states and scattering states.
The internal part, $U^{\rm{int}}$, contains the intrinsic kinetic energy and the potential energy from the interactions between the nucleons within the cluster $V^{\rm{NN}}
(\boldsymbol{k}^{(C)},\boldsymbol{k}^{(C)\prime})$. This term is affected by the Pauli blocking also from the surrounding medium through two mechanisms.
Firstly, the Pauli blocking leads to a modified nucleon-nucleon interaction,
\begin{equation}
\begin{aligned}
\label{potential}
V^{\rm{NN}}(\boldsymbol{k}^{(C)},\boldsymbol{k}^{(C)\prime})= [1-f(\boldsymbol{p}_i)][1-f(\boldsymbol{p}_j)]\Big( V_{\rm{Volkov}}(\boldsymbol{k}^{(C)},\boldsymbol{k}^{(C)\prime})+V_{\rm{Coulomb}}^{(pp)}(\boldsymbol{k}^{(C)},\boldsymbol{k}^{(C)\prime}) \Big).
\end{aligned}
\end{equation}
The operator $f({\boldsymbol{p}_{i}})=\Theta \big( k_f-|\boldsymbol{p}_{i}| \big)$ denotes the occupation of the $i$-th nucleon state below the Fermi momentum $k_f=(3\pi^2 n)^{1/3}$, where $n$ stands for the nucleon density.
The nucleon-nucleon interaction contains the isospin-dependent Volkov interaction \cite{Volkov:1965zz} which follows the form,
\begin{equation}
\begin{aligned}
V_{\rm{Volkov}}(\boldsymbol{k}^{(C)},\boldsymbol{k}^{(C)\prime})= \bigg[ V_a e^{-\mu_a (\boldsymbol{k}^{(C)}-\boldsymbol{k}^{(C)\prime})^2} + V_r e^{-\mu_r (\boldsymbol{k}^{(C)}-\boldsymbol{k}^{(C)\prime})^2} \bigg]  \bigg[ W-M \hat{P}_{\sigma} \hat{P}_{\tau} + B \hat{P}_{\sigma} - H\hat{P}_{\tau} \bigg]
,
\end{aligned}
\end{equation}
and the Coulomb interaction $V_{\rm{Coul}}^{(pp)}(\boldsymbol{k}^{(C)},\boldsymbol{k}^{(C)\prime})$ between two protons. The Coulomb interaction is expressed as a sum of the Gaussian functions in position space and then transformed into momentum space via the Fourier transform.
Secondly, the single-nucleon states below the Fermi surface are blocked out due to the Pauli blocking, leading to a modified intrinsic wave function.
We employ the following trial wave functions expanded in multi-Gaussian bases for 2-, 3- and 4-nucleon clusters,
\begin{equation}
\Psi_{JMTT_z}^{A=2,\rm{intr}}=  \sum_{\beta}  \mathcal{B}_{\beta} \mathcal{A}_{AS} \ \ \Big[ \eta_{\frac{1}{2}}(i)\eta_{\frac{1}{2}}(j) \Big]_{TT_z} \  \Bigg[ \Big[ \chi_{\frac{1}{2}}(i)\chi_{\frac{1}{2}}(j)\Big]_{SS_z}  \ \Phi_{NL}^{A=2}(\boldsymbol{k})
\Bigg]_{JM} \Big[1-f(\boldsymbol{p}_i)\Big]\Big[1-f(\boldsymbol{p}_j)\Big],
\end{equation}
\begin{equation}
\begin{aligned}
\Psi_{JMTT_z}^{A=3,\rm{intr}}=\sum_{\beta}  \sum_{C} \mathcal{B}_{\beta} \mathcal{A}_{AS} \ \ \Big[ [\eta_{\frac{1}{2}}(i)&\eta_{\frac{1}{2}}(j)]_t  \eta_{\frac{1}{2}}(k)\Big]_{TT_z} \  \Bigg[ \Big[ [\chi_{\frac{1}{2}}(i)\chi_{\frac{1}{2}}(j)]_s \chi_{\frac{1}{2}}(k)\Big]_{SS_z}  \ \\
 & \times \Phi_{NL}^{A=3}(\boldsymbol{k}^{(C)},\boldsymbol{K}^{(C)})
\Bigg]_{JM}  \Big[1-f(\boldsymbol{p}_i)\Big]\Big[1-f(\boldsymbol{p}_j)\Big]\Big[1-f(\boldsymbol{p}_k)\Big],
\end{aligned}
\end{equation}
\begin{equation}
\begin{aligned}
\Psi_{JMTT_z}^{A=4,\rm{intr}}=\sum_{\beta}  \sum_{C} \mathcal{B}_{\beta} \mathcal{A}_{AS}& \ \ \Big[ [\eta_{\frac{1}{2}}(i)\eta_{\frac{1}{2}}(j)]_t [\eta_{\frac{1}{2}}(k)\eta_{\frac{1}{2}}(l)]_{\tau}\Big]_{TT_z} \  \Bigg[ \Big[ [\chi_{\frac{1}{2}}(i)\chi_{\frac{1}{2}}(j)]_s [\chi_{\frac{1}{2}}(k)\chi_{\frac{1}{2}}(l)]_{\sigma}\Big]_{SS_z}  \ \\
&\times \Phi_{NL}^{A=4}(\boldsymbol{k}^{(C)},\boldsymbol{K}^{(C)},\boldsymbol{q}^{(C)})
\Bigg]_{JM}
\Big[1-f(\boldsymbol{p}_i)\Big]\Big[1-f(\boldsymbol{p}_j)\Big] \Big[1-f(\boldsymbol{p}_k)\Big]\Big[1-f(\boldsymbol{p}_l)\Big],
\end{aligned}
\end{equation}
\end{widetext}
respectively.
The Pauli blocking effects on the intrinsic wave function are included by employing $[1-f(\boldsymbol{p}_{i(j,k,l)})]$.  $\mathcal{A}_{AS}$ denotes the antisymmetrization operator for the nucleons in the cluster.
$\beta \equiv \{t, \tau, s,\sigma, S, N, L\}$ denotes a specific combination of quantum numbers defining a basis state. The input quantum numbers (excluding the principal quantum number $N$) are summarized in Table~\ref{quantumnumbers}.
$\eta_{\frac{1}{2}}(i)$ and $\chi_{\frac{1}{2}}(i)$ represent the isospin and spin wave functions of the $i$-th nucleon. For 2-nucleon cluster, $(i,j)=(1,2)$. For 3-nucleon cluster, $(i,j,k)=(1,2,3)$, $(2,3,1)$ and $(3,1,2)$ for $C=1-3$, respectively. For 4-nucleon cluster, $(i,j,k,l)=(1,2,3,4)$, $(3,1,2,4)$, $(2,3,1,4)$, $(3,4,1,2)$, $(2,4,3,1)$ and $(1,4,2,3)$ for $C=1-6$, respectively.
The spatial basis wave functions are given by
\begin{equation}
\Phi_{N,L=0}^{A=2}(\boldsymbol{k})=\mathcal{N} e^{-\nu_N k^2}  Y_{00}(\hat{\boldsymbol{k}}),
\end{equation}
\begin{equation}
\Phi_{N,L=0}^{A=3}(\boldsymbol{k},\boldsymbol{K})=\mathcal{N}e^{-\nu_N k^2 -\frac{3}{4}\nu_N K^2 }  Y_{00}(\hat{\boldsymbol{k}}) Y_{00}(\hat{\boldsymbol{K}}),
\end{equation}
\begin{equation}
\begin{aligned}
\Phi_{N,L=0}^{A=4}(\boldsymbol{k},\boldsymbol{K},\boldsymbol{q})=\mathcal{N}&e^{-\nu_N k^2 -\nu_N K^2 -\frac{1}{2}\nu_N q^2}  \\
& \ \ \ \ \times Y_{00}(\hat{\boldsymbol{k}}) Y_{00}(\hat{\boldsymbol{K}}) Y_{00}(\hat{\boldsymbol{q}}),
\end{aligned}
\end{equation}
where $\mathcal{N}$ is for normalization. The Gaussian range parameters $\nu_N$ in the basis functions are chosen to follow a geometric progression,
\begin{equation}
\begin{aligned}
&\nu_N=\omega_N^2/4,
\end{aligned}
\end{equation}
\begin{equation}
\begin{aligned}
\ \ \ \ \ \ \ \ \ \ \ \ \ \ \ \ \ \ \ \ \  &\omega_N=\lambda_1\Big( \frac{\lambda_{\rm{max}}}{\lambda_1} \Big)^{\frac{N-1}{N_{\rm{max}}-1}}.
\end{aligned}
\end{equation}
We set $\{ \lambda_1=0.04 \ {\rm{fm}},\ \lambda_{\rm{max}}=12 \ {\rm{fm}}, \ N_{\rm{max}}=20 \}$ for deuteron and $\{ \lambda_1=0.04 \ {\rm{fm}},\ \lambda_{\rm{max}}=4 \ {\rm{fm}}, \ N_{\rm{max}}=20 \}$ for $\rm{^3H}$, $\rm{^3He}$ and $\alpha$-particle, respectively, which have been chosen to ensure the convergence of our calculation.


\begin{center}
\linespread{1.3}
\begin{table}[ht]
\caption{The input quantum numbers of the basis function for deuteron ($d$), $\rm{^3H}$, $\rm{^3He}$, and $\alpha$-particle.}
\label{quantumnumbers}
\begin{tabular}{p{0.9cm}<{\centering}p{0.8cm}<{\centering}p{0.8cm}<{\centering}p{0.8cm}<{\centering}p{0.8cm}<{\centering}p{0.8cm}<{\centering}p{0.8cm}<{\centering}p{0.8cm}<{\centering}p{0.8cm}
<{\centering} }
\hline\hline
	& $t$ & $\tau$ & $T$ & $T_z$ & $s$ & $\sigma$ & $S$   & $L$      \\
\hline
	$d$ 			&   &  & 0 & 0 &   &  & 1 & 0    	\\
\hline
	$\rm{^3H}$ 	& 0 &  & 1/2 & -1/2 & 1 &  & 1/2 & 0    	\\
					& 1 &  & 1/2 & -1/2 & 0 &  & 1/2 & 0   	\\
\hline
	$\rm{^3He}$ & 0 &  & 1/2 & 1/2 & 1 &  & 1/2 & 0  	\\
					& 1 &  & 1/2 & 1/2 & 0 &  & 1/2 & 0 	\\
\hline
	$\alpha$		& 0 & 0 & 0 & 0 & 1 & 1 & 0 & 0 	\\
					& 1 & 1 & 0 & 0 & 0 & 0 & 0 & 0  	\\
\hline\hline
\end{tabular}
\end{table}
\end{center}

The intrinsic energy $E^{\rm{intr}}$ and the expansion coefficients $\mathcal{B}_{\beta}$ are obtained by solving the in-medium Schr{\"o}dinger equation using the Rayleigh-Ritz variational method.
This requires partitioning the integrals over relative momenta $\boldsymbol{k}^{(C)}$, $\boldsymbol{K}^{(C)}$, and $\boldsymbol{q}^{(C)}$ in the Hamiltonian matrix elements.
For clusters in symmetric nuclear matter, this partition is achieved by introducing an excluded Fermi sphere $k_f=k_p=k_n=(3\pi^2 n_b/2)^{1/3}$ in the integrals, as detailed in Appendix~\ref{snm}.
For an $\alpha$-particle in asymmetric nuclear matter, the proton and neutron Fermi momenta differ, given by $k_p=(3\pi^2 n_p)^{1/3}$ and $k_n=(3\pi^2 n_n)^{1/3}$, respectively.
Accordingly, the integrals of Hamiltonian matrix elements involving neutron-neutron, proton-proton, and proton-neutron pairs must be treated separately. The choice of nucleon pairs is determined by the quantum numbers $t_z$ and $\tau_z$ in the trial wave functions, and the isospin projection onto the corresponding specific states is required. For $\alpha$-particles, this manifests as
\begin{equation}
\label{project}
\begin{aligned}
\langle \hat{H} \rangle = & \langle \hat{P}_{ij}^{(nn)}\Psi_{JMTT_z}^{A=4,{\rm{intr}}} | \ \hat{H} \ | \hat{P}_{ij}^{(nn)}\Psi_{JMTT_z}^{(\prime)A=4,{\rm{intr}}} \rangle \\
+ & \langle \hat{P}_{ij}^{(pp)}\Psi_{JMTT_z}^{A=4,{\rm{intr}}} | \ \hat{H} \ | \hat{P}_{ij}^{(pp)}\Psi_{JMTT_z}^{(\prime)A=4,{\rm{intr}}} \rangle  \\
+&  \langle \hat{P}_{ij}^{(np)}\Psi_{JMTT_z}^{A=4,{\rm{intr}}} | \ \hat{H} \ | \hat{P}_{ij}^{(np)}\Psi_{JMTT_z}^{(\prime)A=4,{\rm{intr}}} \rangle,
\end{aligned}
\end{equation}
where $\hat{P}_{ij}^{(nn)}$, $\hat{P}_{ij}^{(pp)}$ and $\hat{P}_{ij}^{(pn)}$ are the two-body isospin projection operators. Owing to their complexity, the detailed calculations for the integrals in each term of Eq.~(\ref{project})'s right-hand side are elaborated in Appendix \ref{anm}.

\section{Results}
\label{inmedium}
\begin{center}
\linespread{1.3}
\begin{table}[ht]
\caption{The optimized parameters for the Volkov potential.}
\label{parameters}
\begin{tabular}{p{0.1cm}p{2.0cm}<{\centering}  p{2.5cm}<{\centering} p{0.2cm}}
\hline\hline
& Parameter  & Value &       \\
\hline
& $V_r$	& 1.665 $\rm MeV $		       \\
& $\mu_r $ 	& 0.0798 $\rm fm^{2}$		 \\
& $V_a$	& -7.877 $\rm MeV $ 	        \\
& $\mu_a$	& 0.6480 $\rm fm^{2}$		        \\
& $W$ 			& 0.4 			       \\
& $M$ 			& 0.6 			        \\
& $B,H$ 			& 0.07	    \\
\hline\hline
\end{tabular}
\end{table}
\end{center}
\begin{center}
\linespread{1.3}
\begin{table}[ht]
\caption{The intrinsic energies $E$ and the proton point r.m.s. radii $\sqrt{\langle r^2_p\rangle}$ of deuteron ($d$), $\rm{^3H}$, $\rm{^3He}$, and $\alpha$-particle in free space calculated in this work (Cal.) along with their experimental values (Expt.) \cite{ropke2011parametrization}.}
\label{energyinfree}
\begin{tabular}{p{0.7cm}<{\centering}  p{1.7cm}<{\centering} p{1.7cm}<{\centering}p{0.2cm}<{\centering} p{1.7cm}<{\centering} p{1.7cm}<{\centering} }
\hline\hline
   & \multicolumn{2}{c}{ $E \ [\rm{MeV}]$}& & \multicolumn{2}{c}{$\sqrt{\langle r^2_p\rangle} \ [\rm{fm}]$}      \\
\cline{2-3}\cline{5-6}
   & Cal. & Expt. & &Cal. & Expt.      \\
\hline
 $d$	& -2.223 & -2.225 && 2.07 & 1.96 		       \\
 $\rm{^3H}$ 	& -8.484	& -8.482	&& 1.68 & 1.59\\
 $\rm{^3He}$	& -7.779 	&  -7.718 &&  1.85 & 1.76    \\
 $\alpha$	& -28.297	&-28.296 && 1.54 &	1.48        \\
\hline\hline
\end{tabular}
\end{table}
\end{center}

We firstly consider the case of zero-density limit $n_b=0$, $i.e.$, the light clusters in free space. In this case, Eq.~(\ref{schrodinger2}) reduces to the equation of intrinsic motion in free space.
The parameters of the Volkov potential, which are optimized by reproducing the experimental intrinsic energies and proton point root-mean-square (r.m.s.) radii of deuteron, $\rm{^3H}$, $\rm{^3He}$, and $\alpha$-particle in free space, are summarized in Table \ref{parameters}.
The experimental values of proton point r.m.s. radii are taken from Ref. \cite{ropke2011parametrization}, where they are converted from the measured charge radii.
The theoretical proton point r.m.s. radius in our calculation is defined as $\sqrt{\langle r^2_p\rangle}=\left(\int r^2 \rho_p(r) r^2 d r / \int \rho_p(r) r^2 d r\right)^{1 / 2}$,
with the proton density distribution,
\begin{equation}
\begin{aligned}
\label{density}
\rho_p(r)=&\frac{1}{A}\langle \hat{P}_i^{(p)}\Psi_{\rm{position}}|  \\
&\bigg[ \sum_i^{A} \delta\left(r-\left|\boldsymbol{r}_i-\boldsymbol{R}_{\rm{c.m.}}\right|\right)\bigg]|\hat{P}_i^{(p)}\Psi_{\rm{position}}\rangle,
\end{aligned}
\end{equation}
where $\hat{P}_i^{(p)}$ is the one-proton projection operator and $\Psi_{\rm{position}}$ is the intrinsic wave function in position space. $\boldsymbol{r}_i$ and $\boldsymbol{R}_{\rm{c.m.}}$ are the coordinates of the $i$-th nucleon and the center-of-mass of the cluster, respectively.
As shown in Table \ref{energyinfree}, the calculated intrinsic energies and proton point r.m.s. radii are in good agreement with experimental values.
We also calculate the nucleon point r.m.s. radii $\sqrt{\langle r^2\rangle}$ of $\rm{^3H}$ and $\rm{^3He}$ from the nucleon density distribution $\rho(r)$, which is defined analogously to the proton one in Eq.~(\ref{density}) but with the proton projection operator omitted.
The calculated radii for $\rm{^3H}$ and $\rm{^3He}$ are $1.71\ \rm{fm}$ and $1.73\ \rm{fm}$, respectively. The small difference between them arises purely from the Coulomb interaction in our calculation.

\begin{figure}[ht]
\includegraphics[width=8.5cm]{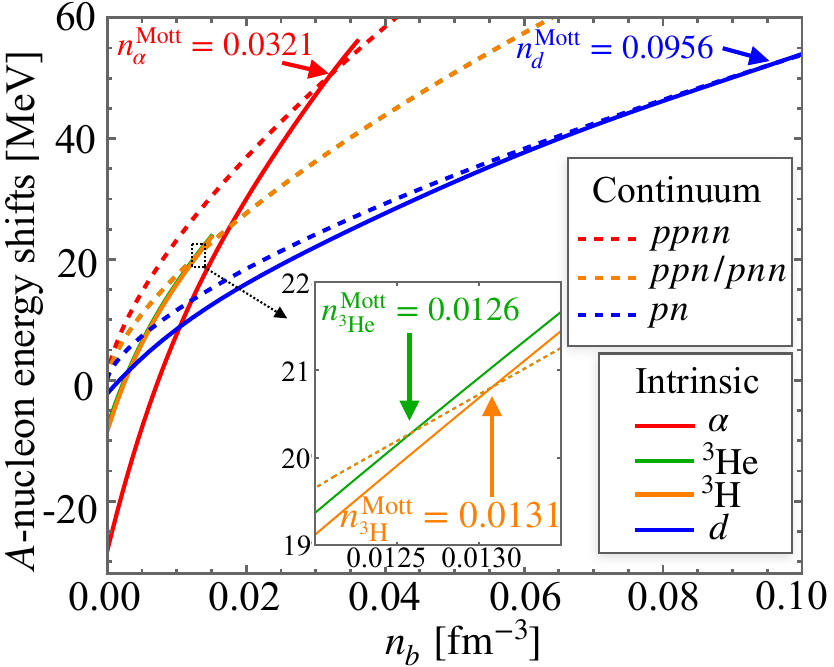}
\caption{\label{symmetryresults} The intrinsic energies (solid curves) of deuteron ($d$), $\rm{^3 H}$, $\rm{^3 He}$ and $\alpha$-particle in isospin symmetric nuclear matter as functions of nucleon density $n_b=n_n+n_p$. The dashed curves denote the continuum edges of $A$ uncorrelated nucleons ($Z$ protons and $N$ neutrons). Note that $\rm{^3 H}$ and $\rm{^3 He}$ share the same continuum edge in symmetric nuclear matter. }
\end{figure}

We next turn to the case of light clusters embedded in isospin symmetric nuclear matter. Fig.~\ref{symmetryresults} shows the intrinsic energies of deuteron, $\rm{^3 H}$, $\rm{^3 He}$ and $\alpha$-particle in symmetric nuclear matter as a function of density $n_b$. The edges for the continuum of scattering states, which are defined as
\begin{equation}
\label{continuumedge}
\begin{aligned}
E_{c}=&Z E_{{\rm{Fermi}}}(n_p)+N E_{{\rm{Fermi}}}(n_n)\\
=& Z\frac{\hbar^2}{2m}\big( 3\pi^2 n_p \big)^{\frac{2}{3}}+N\frac{\hbar^2}{2m}\big( 3\pi^2 n_n \big)^{\frac{2}{3}},
\end{aligned}
\end{equation}
are also given in Fig.~\ref{symmetryresults}. Only the kinetic energy of the $A$ uncorrelated nucleons ($Z$ protons and $N$ neutrons) is needed to determine continuum thresholds. This is because the external energy shifts $U^{\rm{ext}}$ for the scattering states and bound states cancel out in the comparison between the continuum threshold $E_{c}$ and the intrinsic energy $E^{\rm{intr}}$. In Fig.~\ref{symmetryresults} and the following figures, we neglect $U^{\rm{ext}}$ in all comparisons.

As illustrated in Fig.~\ref{symmetryresults},  the light clusters are well bound at the zero-density limit. With increasing medium density $n_b$, however, their intrinsic energies increase rapidly due to the antisymmetrization and Pauli blockings. A bound state exists for the $A$-nucleon system as long as its intrinsic energy lies below the corresponding continuum edge ($i.e.$, $n_b < n^{\rm{Mott}}$). The Mott transition occurs at the density $n^{\rm{Mott}}$ where the intrinsic energy curve intersects the continuum edge. Beyond this critical density ($n_b > n^{\rm{Mott}}$), the cluster dissolves and its constituent nucleons turn into uncorrelated nucleons added on top of the Fermi surface.

\begin{figure}[ht]
\includegraphics[width=8.5cm]{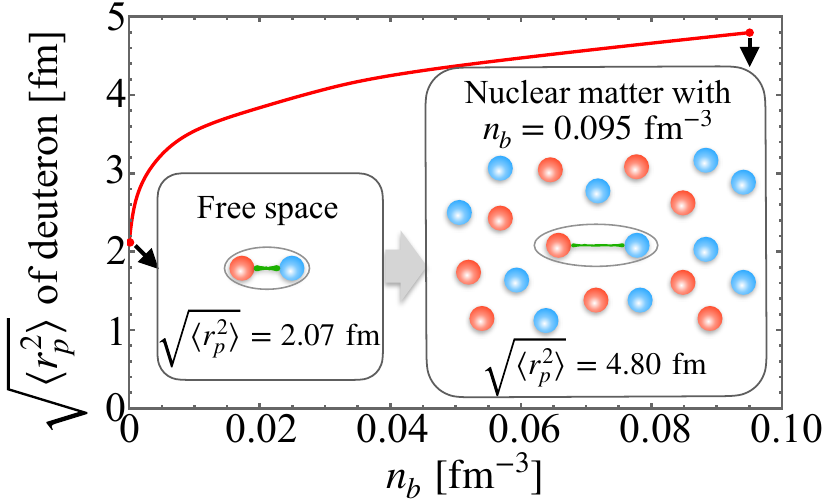}
\caption{\label{fig-rms} The proton point r.m.s. radius $\sqrt{\langle r^2_p \rangle}$ of deuteron as a function of the nucleon density $n_b$ in isospin symmetric nuclear matter. The insets provide schematic illustrations of the deuteron in free space and when embedded in nuclear matter.}
\end{figure}

\begin{table}[ht]
\renewcommand{\arraystretch}{1.3}
\begin{center}
\caption{The first-order Pauli blocking shift $\Delta^{\rm{Pauli}}$ (in $\rm{MeV\ fm^{3}}$) of deuteron ($d$), $\rm{^3H}$, $\rm{^3He}$, and $\alpha$-particle in symmetric nuclear matter at temperature $T$ (in $\rm{MeV}$).}
\label{slope}
\begin{tabular}{p{1.2cm}<{\centering} p{1.2cm}<{\centering}p{1.2cm}<{\centering} p{1.2cm}<{\centering} p{0.5cm}<{\centering}p{2.6cm}<{\centering}  }
\hline\hline
    $\Delta^{\rm{Pauli}}_d$ & $\Delta^{\rm{Pauli}}_{\rm{^3H}}$  & $\Delta^{\rm{Pauli}}_{\rm{^3He}}$ &  $\Delta^{\rm{Pauli}}_{\alpha}$ & $T$ & $\rm{Ref.}$ 		       \\
    \hline
 	 1161.13	& 3682.32	& 3501.23 & 4733.94 & 0 & This work \\
  	 1756.9	& 2747.92	& 3038.96 & 4434.30 & 1 & Jastrow  \cite{Ropke:2008qk} \\
  	 2169.8	& 2808.9	& 3104.5 & 4123.0 & 1 & Gaussian \cite{Ropke:2008qk}  \\
\hline\hline
\end{tabular}
\end{center}
\end{table}

We extract the first-order Pauli blocking shift $\Delta^{\rm{Pauli}}$ for different light clusters in symmetric nuclear matter, $i.e.$, the slope of the intrinsic energy shift at the zero-density limit (see Fig.~\ref{symmetryresults}). As listed in Table \ref{slope}, the $\Delta^{\rm{Pauli}}$ follows a sequence: $d<{\rm{^3H}/\rm{^3He}}<\alpha$. This is understandable as the in-medium effect in principle increases with the number of nucleons inside the cluster.
This sequence is consistent with the results in Ref.\cite{Ropke:2008qk} (see Table~\ref{slope}) where the in-medium Schr{\"o}dinger equation is solved using both the Gaussian and Jastrow wave functions. Note that the finite-temperature $T$ is also considered in Ref.\cite{Ropke:2008qk} where the thermal effect is found to reduce the strong Pauli blockings.
The Mott density, however, is determined by not only the strength of these in-medium effects but also the binding energies of each cluster in free space. For the $\alpha$-particle in symmetric nuclear matter, the refined Mott density is $n_{\alpha}^{\rm{Mott}}=0.0321 \ \rm{fm^{-3}}$, corresponding to approximately $0.2 \ n_0$, where $n_0$ is the nuclear matter saturation density. This value is consistent with the previous result \cite{Ropke:2014wsa}.
The Mott densities of the $\rm{^3 He}$ and $\rm{^3 H}$ are lower than that of the $\alpha$-particle, $n_{\rm{^3 He}}^{\rm{Mott}}=0.0126 \ \rm{fm^{-3}}$ and $n_{\rm{^3 H}}^{\rm{Mott}}=0.0131 \ \rm{fm^{-3}}$. The small difference between $\rm{^3 He}$ and $\rm{^3 H}$ is mainly due to the Coulomb interaction.
The deuteron, in contrast, has a significantly larger Mott density in symmetric nuclear matter, $n_{d}^{\rm{Mott}}=0.0956 \ \rm{fm^{-3}}$. By calculating the proton point r.m.s. radius $\sqrt{\langle r^2_p \rangle}$ of the deuteron, we found that the spatial structure of the deuteron becomes increasingly extended with density, as illustrated in Fig.~\ref{fig-rms}. Prior to the Mott transition, the r.m.s radius expands to about  $4.80\ \rm fm$, roughly 2.3 times its free-space value. This finding suggests that the weakly bound deuteron may survive from low to relatively high densities in symmetric matter.

It is noteworthy that the variational approach may introduce a sharp transition from the bound state to the continuum. Refinements of the present approach include adopting alternative basis functions, such as the Lorentzian-type \cite{Ropke:2008qk,ropke2011parametrization} for deuteron, and incorporating two-nucleon correlations that  may persist in the continuum after the Mott transition. Exact solution of few-body system embedded in medium should be tackled in future, where the bound state is expected to merge asymptotically with the continuum as density increases.

\begin{figure}[ht]
\includegraphics[width=8.5cm]{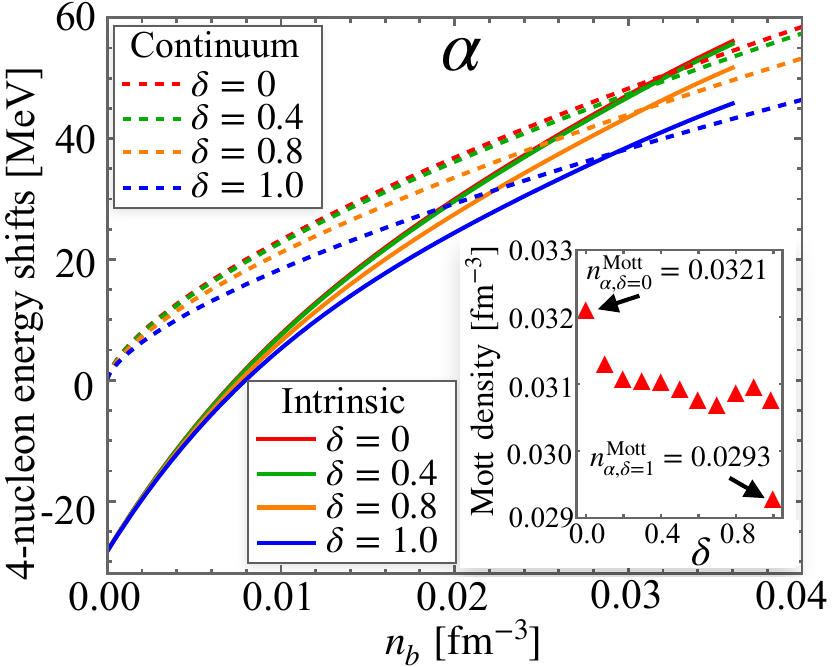}
\caption{\label{fig-asy-alpha} The intrinsic energy of $\alpha$-particle in nuclear matter with isospin asymmetry $\delta$ as a function of density $n_b=n_n+n_p$ (solid curves). The dashed curves denote the continuum edges of four uncorrelated nucleons (two protons and two neutrons). The inset illustrates the dependence of the Mott density on the isospin asymmetry $\delta$.}
\end{figure}

Finally, we examine the behavior of the $\alpha$-particle in asymmetric nuclear matter. Fig.~\ref{fig-asy-alpha} displays its intrinsic energy as a function of baryon density for varying isospin asymmetry $\delta$. An increase in $\delta$ leads to a growing separation of neutron and proton Fermi surfaces. At a fixed density $n_b$, this results in a non-linear decrease of both the intrinsic energy and the continuum edge. Consequently,  the Mott density exhibits a non-monotonic dependence on $\delta$, as shown in the inset of Fig.~\ref{fig-asy-alpha}. Specifically, the Mott density of $\alpha$-particle in pure neutron matter, $i.e.$, $\delta=1$, is $n_{\alpha,\delta=1}^{\rm{Mott}}=0.0293 \ \rm{fm^{-3}}$, slightly lower than that in the symmetric matter. This finding is relevant for understanding $\alpha$-clustering in neutron-rich environments and the neutron skin thickness of finite nuclei.

\section{Summary}
\label{summary}
The description of formation and dissolution of various light clusters in nuclear medium poses challenges for microscopic studies.
A particular difficulty lies in accurately treating the Pauli blocking effect within a unified framework capable of describing different clusters.
In this work, we have refined a  microscopic in-medium few-body approach in momentum space using multi-Gaussian basis functions, which allows for a precise formulation of the wave function modifications induced by Pauli blocking. Our results reveal a complex picture of the formation and dissolution of various light clusters in nuclear medium. It is found that the loosely bound deuteron survives to higher densities as compared with $\rm ^3H$, $\rm ^3He$ and $\alpha$-particles and exhibits significant spatial expansion before dissolution, analogous to BCS-type $n$-$p$ pairing in $^3SD_1$ channel \cite{Elgaroy:1998zz,Fan:2019cij,Yin:2023uue,Ropke:1998qs}.
Furthermore, the Mott density for $\alpha$-particle in neutron-rich matter is slightly lower than that in symmetric matter, suggesting that the $\alpha$-particle may be formed even on the surface of nuclei with a thick neutron skin.

Future extensions of the present study will involve adopting more realistic nucleon-nucleon interactions and three-body forces, as well as more refined few-body wave functions. It would also be important to consider finite temperatures ($T\ne 0$) \cite{Ropke:1982ino} and finite center-of-mass momenta ($\boldsymbol{P}\ne 0$) \cite{Ropke:2008qk,ropke2011parametrization} in our calculations. These extensions will provide more comprehensive insights into cluster formation and dissolution in intermediate-energy heavy-ion collisions and neutron star crusts.

\section*{Acknowledgement}
Discussions with G. R\"opke, R. J. Li and Y. Wu are gratefully acknowledged.
This work is supported by the National Natural Science
Foundation of China (Grant No. 12275129).

\begin{widetext}
\appendix

\section{Calculation of Matrix Elements}
\label{appendix}

We present the detailed integrals for the in-medium Hamiltonian matrix elements, considering only clusters with zero total momentum $\boldsymbol{P}=0$. The core integrand is represented by the function $\mathcal{F}$. Its functional form depends on the cluster type and the matrix element.
For normalization and kinetic energy matrix elements,  it is
$\mathcal{F}(\boldsymbol{k})$ (deuteron), $\mathcal{F}(\boldsymbol{k}, \boldsymbol{K})$ ($\rm{^3He}$/$\rm{^3H}$), or $\mathcal{F}(\boldsymbol{k}, \boldsymbol{K}, \boldsymbol{q})$ ($\alpha$-particle). For potential energy matrix elements, an additional variable $\boldsymbol{k}'$ appears in each  $\mathcal{F}$.
The following sections detail these integrals for deuteron and $\rm{^3He}$/$\rm{^3H}$ in symmetric nuclear matter and $\alpha$-particle in both symmetric and asymmetric nuclear matter.

\subsection{Symmetric nuclear matter}
\label{snm}

In symmetric nuclear matter, neutrons and protons constituting the light clusters feel the same Pauli blocking, where the single-nucleon states below the Fermi surface $k_f=k_n=k_p$ are blocked out.

\subsubsection{Deuteron}

For deuteron, the momentum of nucleons 1 and 2 obey the conditions,
\begin{equation}
\begin{aligned}
|\boldsymbol{p}_1|&=|-\boldsymbol{k}|\ge k_{f},  \\
|\boldsymbol{p}_2|&=|\boldsymbol{k}|\ge k_{f}.  \\
\end{aligned}
\end{equation}
The integral of $\mathcal{F}(\boldsymbol{k})$ is
\begin{equation}
\label{2sym}
\begin{aligned}
\int  d\boldsymbol{k} \mathcal{F}(\boldsymbol{k}) =4\pi \int_{k_f}^{\infty} k^2 dk \mathcal{F}(k),
\end{aligned}
\end{equation}
where the integrand is $\mathcal{F}(k)=\phi_i^*(k)\phi_j(k)$ for the normalization matrix element, and $\phi_i^*(k) \hat{T}_{2N} \phi_j(k)$ for the kinetic energy matrix element. Here, $\hat{T}_{2N}$ denotes the kinetic energy operator. For the potential energy matrix element, the integrand becomes $\mathcal{F}(k,k')=\phi_i^*(k)\hat{V}\phi_j(k')$. $\hat{V}$ is the potential energy operator, and $\phi_{i(j)}(k^{(\prime)})$ refers to  the $i(j)$-th basis function.

\subsubsection{$\rm{^3H}$ and $\rm{^3He}$}

For $\rm{^3H}$ and $\rm{^3He}$, the wave function is written in 3 rearrangement channels of Jacobi momenta.
We take the case of $C=1$ in Fig.~\ref{jacobiK3} as an example. The bra and ket wave functions of all rearrangement channels are transformed into $\phi_i(\boldsymbol{k}^{(1)},\boldsymbol{K}^{(1)})$ and $\phi_j(\boldsymbol{k}^{(1)},\boldsymbol{K}^{(1)})$, respectively. In the following, we omit the superscripts in $\boldsymbol{k}^{(1)}$ and $\boldsymbol{K}^{(1)}$.
The momenta of nucleons 1, 2 and 3 obey the conditions,
\begin{equation}
\begin{aligned}
|\boldsymbol{p}_1|&=|-\boldsymbol{K}/2-\boldsymbol{k}|\ge k_{f},  \\
|\boldsymbol{p}_2|&=|-\boldsymbol{K}/2+\boldsymbol{k}|\ge k_{f},  \\
|\boldsymbol{p}_3|&=|\boldsymbol{K}|\ge k_{f}.  \\
\end{aligned}
\end{equation}
The integral of $\mathcal{F}(\boldsymbol{k},\boldsymbol{K})$ over $\boldsymbol{K}$ is divided into two parts:
\begin{equation}
\label{he3intf}
\begin{aligned}
\int d\boldsymbol{k}d\boldsymbol{K} \mathcal{F}(\boldsymbol{k},\boldsymbol{K})=4\pi \Bigg[\int_{k_f}^{2k_f} K^2dK G^{(1)}(K) + \int_{2k_f}^{\infty} K^2dK G^{(2)}(K)\Bigg].
\end{aligned}
\end{equation}
We fix the direction of $\boldsymbol{K}$ and take $z={\rm{cos}}(\boldsymbol{k},\boldsymbol{K})$. Then we have
\begin{equation}
\label{3sym1}
\begin{aligned}
G^{(1)}(K)=&\int d\boldsymbol{k} \mathcal{F}(k,z,K)	\\
=& 2\pi \Bigg[\int_{\sqrt{k_f^2-\frac{K^2}{4}}}^{k_f+\frac{K}{2}} k^2 dk \int_{\frac{(k_f^2-\frac{K^2}{4}-k^2)}{Kk}}^{\frac{(-k_f^2+\frac{K^2}{4}+k^2)}{Kk}} d z \mathcal{F}(k,z,K) + \int_{k_f+\frac{K}{2}}^{\infty} k^2 dk \int_{-1}^1 d z \mathcal{F}(k,z,K)\Bigg],
\end{aligned}
\end{equation}
\begin{equation}
\label{3sym2}
\begin{aligned}
G^{(2)}(K)=&\int d\boldsymbol{k} \mathcal{F}(k,z,K)	\\
=& 2\pi \Bigg[ \int_{0}^{-k_f+\frac{K}{2}}  k^2 dk \int_{-1}^1 d z \mathcal{F}(k,z,K)   + \int_{-k_f+\frac{K}{2}}^{k_f+\frac{K}{2}} k^2 dk \int_{\frac{(k_f^2-\frac{K^2}{4}-k^2)}{Kk}}^{\frac{(-k_f^2+\frac{K^2}{4}+k^2)}{Kk}}  d z \mathcal{F}(k,z,K)  \\
&+ \int_{k_f+\frac{K}{2}}^{\infty} k^2 dk \int_{-1}^1 d z \mathcal{F}(k,z,K)\Bigg],
\end{aligned}
\end{equation}
where the integrand for the normalization and kinetic energy matrix elements are $\mathcal{F}(k,z,K)=\phi_i^*(k,z,K)\phi_j(k,z,K)$ and $\phi_i^*(k,z,K)\hat{T}_{3N}\phi_j(k,z,K)$, respectively. $\hat{T}_{3N}$ is the kinetic energy operator for $\rm{^3H}$ or $\rm{^3He}$.
For the potential energy matrix element, a 5-fold integral is required and the corresponding integrand is $\mathcal{F}(k,z,k',z',K)=\phi_i^*(k,z,K)\hat{V}\phi_j(k',z',K)$ with $z'={\rm{cos}}(\boldsymbol{k}',\boldsymbol{K})$.

\subsubsection{$\alpha$-particle}
\label{app-alpha}
For $\alpha$-particle, the wave function is written in 6 rearrangement channels of Jacobi momenta.
We take the case of $C=1$ in Fig.~\ref{jacobiK4} as an example. Similarly, the bra and ket wave functions of all rearrangement channels are transformed into $\phi_i(\boldsymbol{k}^{(1)},\boldsymbol{K}^{(1)},\boldsymbol{q}^{(1)})$ and $\phi_j(\boldsymbol{k}^{(1)},\boldsymbol{K}^{(1)},\boldsymbol{q}^{(1)})$, respectively.
The momenta of nucleons 1, 2, 3 and 4 obey the conditions,
\begin{equation}
\begin{aligned}
|\boldsymbol{p}_1|&=|\boldsymbol{q}/2-\boldsymbol{k}|\ge k_f, \\
|\boldsymbol{p}_2|&=|\boldsymbol{q}/2+\boldsymbol{k}|\ge k_f, \\
|\boldsymbol{p}_3|&=|-\boldsymbol{q}/2-\boldsymbol{K}|\ge k_f, \\
|\boldsymbol{p}_4|&=|-\boldsymbol{q}/2+\boldsymbol{K}|\ge k_f. \\
\end{aligned}
\end{equation}
The integral of $\mathcal{F}(\boldsymbol{k},\boldsymbol{K},\boldsymbol{q})$ over $\boldsymbol{q}$ is divided into two parts:
\begin{equation}
\begin{aligned}
\int d\boldsymbol{k}d\boldsymbol{K}d\boldsymbol{q} \mathcal{F}(\boldsymbol{k},\boldsymbol{K},\boldsymbol{q})=4\pi \Bigg[ \int_0^{2k_f} q^2dq G^{(1)}(q) +\int_{2k_f}^{\infty} q^2dq G^{(2)}(q)\Bigg].
\end{aligned}
\end{equation}
We fix the direction of $\boldsymbol{q}$ and take $z={\rm{cos}}(\boldsymbol{k},\boldsymbol{q})$ and $Z={\rm{cos}}(\boldsymbol{K},\boldsymbol{q})$. We have
\begin{equation}
\begin{aligned}
G^{(1)}(q)=&\int d\boldsymbol{k} H^{(1)}(k,z,q)	\\
=& 2\pi \Bigg[\int_{\sqrt{k_f^2-\frac{q^2}{4}}}^{k_f+\frac{q}{2}} k^2 dk \int_{\frac{(k_f^2-\frac{q^2}{4}-k^2)}{qk}}^{\frac{(-k_f^2+\frac{q^2}{4}+k^2)}{qk}} d z H^{(1)}(k,z,q) +  \int_{k_f+\frac{q}{2}}^{\infty} k^2 dk \int_{-1}^1 d z H^{(1)}(k,z,q) \Bigg],
\end{aligned}
\end{equation}
\begin{equation}
\begin{aligned}
G^{(2)}(q)=&\int d\boldsymbol{k} H^{(2)}(k,z,q)	\\
=& 2\pi \Bigg[ \int_{0}^{-k_f+\frac{q}{2}}  k^2 dk \int_{-1}^1 d z H^{(2)}(k,z,q)  + \int_{-k_f+\frac{q}{2}}^{k_f+\frac{q}{2}} k^2 dk \int_{\frac{(k_f^2-\frac{q^2}{4}-k^2)}{qk}}^{\frac{(-k_f^2+\frac{q^2}{4}+k^2)}{qk}}  d z H^{(2)}(k,z,q) \\
& + \int_{k_f+\frac{q}{2}}^{\infty} k^2 dk \int_{-1}^1 d z H^{(2)}(k,z,q) \Bigg],
\end{aligned}
\end{equation}
\begin{equation}
\begin{aligned}
H^{(1)}(k,z,q)=&\int d\boldsymbol{K} \mathcal{F}(k,z,K,Z,q)	\\
=& 2\pi \Bigg[\int_{\sqrt{k_f^2-\frac{q^2}{4}}}^{k_f+\frac{q}{2}} K^2 dK \int_{\frac{(k_f^2-\frac{q^2}{4}-K^2)}{qK}}^{\frac{(-k_f^2+\frac{q^2}{4}+K^2)}{qK}} d Z \mathcal{F}(k,z,K,Z,q) +  \int_{k_f+\frac{q}{2}}^{\infty} K^2 dK \int_{-1}^1 d Z \mathcal{F}(k,z,K,Z,q) \Bigg],
\end{aligned}
\end{equation}
\begin{equation}
\begin{aligned}
H^{(2)}(k,z,q)=&\int d\boldsymbol{K} \mathcal{F}(k,z,K,Z,q)	\\
=& 2\pi \Bigg[ \int_{0}^{-k_f+\frac{q}{2}}  K^2 dK \int_{-1}^1 d Z \mathcal{F}(k,z,K,Z,q)  + \int_{-k_f+\frac{q}{2}}^{k_f+\frac{q}{2}} K^2 dK \int_{\frac{(k_f^2-\frac{q^2}{4}-K^2)}{qK}}^{\frac{(-k_f^2+\frac{q^2}{4}+K^2)}{qK}}  d Z \mathcal{F}(k,z,K,Z,q) \\
& + \int_{k_f+\frac{q}{2}}^{\infty} K^2 dK \int_{-1}^1 d Z \mathcal{F}(k,z,K,Z,q) \Bigg].
\end{aligned}
\end{equation}
For the normalization matrix element, the integrand is $\mathcal{F}(k,z,K,Z,q)=\phi_i^*(k,z,K,Z,q)\phi_j(k,z,K,Z,q)$. The integrand for the kinetic energy matrix element is $\phi_i^*(k,z,K,Z,q)\hat{T}_{4N}\phi_j(k,z,K,Z,q)$, in which $\hat{T}_{4N}$ is the kinetic energy operator for $\alpha$-particle.
For the potential energy matrix element, a 7-fold integral of the integrand $\mathcal{F}(k,z,k',z',K,Z,q)=\phi_i^*(k,z,K,Z,q)\hat{V}\phi_j(k',z',K,Z,q)$ is required. Note that $z'={\rm{cos}}(\boldsymbol{k}',\boldsymbol{q})$.

\subsection{Asymmetric nuclear matter}
\label{anm}
In asymmetric nuclear matter, neutrons and protons constituting the light clusters feel different Pauli blockings.
Our focus here is on the $\alpha$-particle in asymmetric nuclear matter. Similar to the case of $\alpha$-particle in symmetric matter, we also take the case of $C=1$ in Fig.~\ref{jacobiK4} as an example. The momenta of nucleons 1, 2, 3 and 4 obey the conditions,
\begin{equation}
\begin{aligned}
|\boldsymbol{p}_1|&=|\boldsymbol{q}/2-\boldsymbol{k}|\ge k_{n(p)}, \\
|\boldsymbol{p}_2|&=|\boldsymbol{q}/2+\boldsymbol{k}|\ge k_{n(p)}, \\
|\boldsymbol{p}_3|&=|-\boldsymbol{q}/2-\boldsymbol{K}|\ge k_{n(p)}, \\
|\boldsymbol{p}_4|&=|-\boldsymbol{q}/2+\boldsymbol{K}|\ge k_{n(p)}, \\
\end{aligned}
\end{equation}
where the choice of $k_{n}$ or $k_{p}$ depends on $t_z$ and $\tau_z$.

In the case of $(t_z,\tau_z)=(1,-1)$, $i.e.$, nucleon pair 1-2 being a proton-proton pair, the integral of $\mathcal{F}_{pp}(\boldsymbol{k},\boldsymbol{K},\boldsymbol{q})$ over $\boldsymbol{q}$ is divided into two parts:
\begin{equation}
\begin{aligned}
\int d\boldsymbol{k}d\boldsymbol{K}d\boldsymbol{q} \mathcal{F}_{pp}(\boldsymbol{k},\boldsymbol{K},\boldsymbol{q})=4\pi \Bigg[ \int_0^{2k_{p}} q^2dq G_{pp}^{(1)}(q) +\int_{2k_{p}}^{\infty} q^2dq G_{pp}^{(2)}(q)\Bigg].
\end{aligned}
\end{equation}
We fix the direction of $\boldsymbol{q}$ and take $z={\rm{cos}}(\boldsymbol{k},\boldsymbol{q})$ and $Z={\rm{cos}}(\boldsymbol{K},\boldsymbol{q})$. Then we have
\begin{equation}
\begin{aligned}
G_{pp}^{(1)}(q)=&\int d\boldsymbol{k} H^{(1)}_{pp}(k,z,q)	\\
=& 2\pi \Bigg[\int_{\sqrt{k_{p}^2-\frac{q^2}{4}}}^{k_{p}+\frac{q}{2}} k^2 dk \int_{\frac{(k_{p}^2-\frac{q^2}{4}-k^2)}{qk}}^{\frac{(-k_{p}^2+\frac{q^2}{4}+k^2)}{qk}} d z H^{(1)}_{pp}(k,z,q) +  \int_{k_{p}+\frac{q}{2}}^{\infty} k^2 dk \int_{-1}^1 d z H^{(1)}_{pp}(k,z,q) \Bigg],
\end{aligned}
\end{equation}
\begin{equation}
\begin{aligned}
G_{pp}^{(2)}(q)=&\int d\boldsymbol{k} H^{(2)}_{pp}(k,z,q)	\\
=& 2\pi \Bigg[ \int_{0}^{-k_p+\frac{q}{2}}  k^2 dk \int_{-1}^1 d z H^{(2)}_{pp}(k,z, q)  + \int_{-k_p+\frac{q}{2}}^{k_p+\frac{q}{2}} k^2 dk \int_{\frac{(k_p^2-\frac{q^2}{4}-k^2)}{qk}}^{\frac{(-k_p^2+\frac{q^2}{4}+k^2)}{qk}}  d z H^{(2)}_{pp}(k,z,q) \\
& + \int_{k_p+\frac{q}{2}}^{\infty} k^2 dk \int_{-1}^1 d z H^{(2)}_{pp}(k,z,q) \Bigg],
\end{aligned}
\end{equation}
\begin{equation}
\begin{aligned}
H_{pp}^{(1)}(k,z,q)=&\int d\boldsymbol{K} \mathcal{F}_{pp}(k,z,K,Z,q)	\\
=& 2\pi \Bigg[\int_{\sqrt{k_{p}^2-\frac{q^2}{4}}}^{k_{p}+\frac{q}{2}} K^2 dK \int_{\frac{(k_{p}^2-\frac{q^2}{4}-K^2)}{qK}}^{\frac{(-k_{p}^2+\frac{q^2}{4}+K^2)}{qK}} d Z \mathcal{F}_{pp}(k,z,K,Z,q) +  \int_{k_{p}+\frac{q}{2}}^{\infty} K^2 dK \int_{-1}^1 d Z \mathcal{F}_{pp}(k,z,K,Z,q) \Bigg],
\end{aligned}
\end{equation}
\begin{equation}
\begin{aligned}
H_{pp}^{(2)}(k,z,q)=&\int d\boldsymbol{K} \mathcal{F}_{pp}(k,z,K,Z,q)	\\
=& 2\pi \Bigg[ \int_{0}^{-k_p+\frac{q}{2}}  K^2 dK \int_{-1}^1 d Z \mathcal{F}_{pp}(k,z,K,Z,q)  + \int_{-k_p+\frac{q}{2}}^{k_p+\frac{q}{2}} K^2 dK \int_{\frac{(k_p^2-\frac{q^2}{4}-K^2)}{qK}}^{\frac{(-k_p^2+\frac{q^2}{4}+K^2)}{qK}}  d Z \mathcal{F}_{pp}(k,z,K,Z,q) \\
& + \int_{k_p+\frac{q}{2}}^{\infty} K^2 dK \int_{-1}^1 d Z \mathcal{F}_{pp}(k,z,K,Z,q) \Bigg].
\end{aligned}
\end{equation}
The integrand $\mathcal{F}_{pp}(k,z,K,Z,q)$ for the normalization matrix element is $\hat{P}_{12}^{(pp)}\phi^{*}_i(k,z,K,Z,q)\hat{P}_{12}^{(pp)}\phi_j(k,z,K,Z,q)$ and for the kinetic energy matrix element is $\hat{P}_{12}^{(pp)}\phi_i^*(k,z,K,Z,q)\hat{T}_{4N}\hat{P}_{12}^{(pp)}\phi_j(k,z,K,Z,q)$. For the potential energy matrix element, the integrand is $\mathcal{F}_{pp}(k,z,k',z',K,Z,q)=\hat{P}_{12}^{(pp)}\phi_i^*(k,z,K,Z,q)\hat{V}\hat{P}_{12}^{(pp)}\phi_j(k',z',K,Z,q)$ with $z'={\rm{cos}}(\boldsymbol{k}',\boldsymbol{q})$.
In the case of $(t_z,\tau_z)=(-1,1)$, $i.e.$, nucleon pair 1-2 being a neutron-neutron pair, the integral of $\mathcal{F}_{nn}(\boldsymbol{k},\boldsymbol{K},\boldsymbol{q})$ is similar to the case of $(t_z,\tau_z)=(1,-1)$, but with $k_p$ replaced by $k_n$.
In the case of $(t_z,\tau_z)=(0,0)$, $i.e.$, nucleon pair 1-2 being a neutron-proton pair, the integral of $\mathcal{F}_{np}(\boldsymbol{k},\boldsymbol{K},\boldsymbol{q})$ over $\boldsymbol{q}$ is divided into four parts:

\begin{equation}
\begin{aligned}
\int d\boldsymbol{k}d\boldsymbol{K}d\boldsymbol{q} \mathcal{F}_{np}(\boldsymbol{k},\boldsymbol{K},\boldsymbol{q})=& 4\pi \Bigg[ \int_0^{k_{n}-k_{p}} q^2dq G_{np}^{(1)}(q) + \int_{k_{n}-k_{p}}^{k_{n}+k_{p}} q^2dq G_{np}^{(2)}(q) \\
& + \int_{k_{n}+k_{p}}^{2k_{n}} q^2dq G_{np}^{(3)}(q) +\int_{2k_{n}}^{\infty} q^2dq G_{np}^{(4)}(q) \Bigg],
\end{aligned}
\end{equation}
where
\begin{equation}
\begin{aligned}
G_{np}^{(1)}(q)=&\int d\boldsymbol{k} H^{(1)}_{np}(k,z,q)	\\
=& 2\pi \Bigg[\int_{k_{n}-\frac{q}{2}}^{k_{n}+\frac{q}{2}}  k^2 dk \int_\frac{(k_{n}^2-\frac{q^2}{4}-k^2)}{qk}^1 d z H^{(1)}_{np}(k,z,q)  + \int_{k_{n}+\frac{q}{2}}^\infty   k^2 dk \int_{-1}^1 d z H^{(1)}_{np}(k,z,q) \Bigg],
\end{aligned}
\end{equation}
\begin{equation}
\begin{aligned}
G_{np}^{(2)}(q)=&\int d\boldsymbol{k} H^{(2)}_{np}(k,z,q)	\\
=& 2\pi \Bigg[ \int_{\frac{1}{2}\sqrt{2k_{n}^2+2k_{p}^2-q^2}}^{k_{p}+\frac{q}{2}}  k^2 dk \int_\frac{(k_{n}^2-\frac{q^2}{4}-k^2)}{qk}^\frac{(-k_{p}^2+\frac{q^2}{4}+k^2)}{qk} d z H^{(2)}_{np}(k,z,q) \\
&+ \int_{k_{p}+\frac{q}{2}}^{k_{n}+\frac{q}{2}} k^2 dk \int_\frac{(k_{n}^2-\frac{q^2}{4}-k^2)}{qk}^1 d z H^{(2)}_{np}(k,z,q)
+ \int_{k_{n}+\frac{q}{2}}^\infty k^2 dk \int_{-1}^1 d z H^{(2)}_{np}(k,z,q) \Bigg],
\end{aligned}
\end{equation}
\begin{equation}
\begin{aligned}
G_{np}^{(3)}(q)=&\int d\boldsymbol{k} H^{(3)}_{np}(k,z,q)	\\
=& 2\pi \Bigg[ \int_{k_{n}-\frac{q}{2}}^{-k_{p}+\frac{q}{2}}  k^2 dk \int_\frac{(k_{n}^2-\frac{q^2}{4}-k^2)}{qk}^1 d z H^{(3)}_{np}(k,z,q)
+ \int_{-k_{p}+\frac{q}{2}}^{k_{p}+\frac{q}{2}} k^2 dk \int_\frac{(k_{n}^2-\frac{q^2}{4}-k^2)}{qk}^\frac{(-k_{p}^2+\frac{q^2}{4}+k^2)}{qk}  d z H^{(3)}_{np}(k,z,q)  \\
 & + \int_{k_{p}+\frac{q}{2}}^{k_{n}+\frac{q}{2}} k^2 dk \int_\frac{(k_{n}^2-\frac{q^2}{4}-k^2)}{qk}^{1} d z H^{(3)}_{np}(k,z,q)
+ \int_{k_{n}+\frac{q}{2}}^{\infty} k^2 dk \int_{-1}^1 d z H^{(3)}_{np}(k,z,q) \Bigg],
\end{aligned}
\end{equation}
\begin{equation}
\begin{aligned}
G_{np}^{(4)}(q)=&\int d\boldsymbol{k} H^{(4)}_{np}(k,z,q)	\\
=& 2\pi \Bigg[ \int_{0}^{-k_{n}+\frac{q}{2}}  k^2 dk \int_{-1}^1 d z H^{(4)}_{np}(k,z,q)
+ \int_{-k_{n}+\frac{q}{2}}^{-k_{p}+\frac{q}{2}}  k^2 dk \int_\frac{(k_{n}^2-\frac{q^2}{4}-k^2)}{qk}^1 d z H^{(4)}_{np}(k,z,q)  \\
& + \int_{-k_{p}+\frac{q}{2}}^{k_{p}+\frac{q}{2}} k^2 dk \int_\frac{(k_{n}^2-\frac{q^2}{4}-k^2)}{qk}^\frac{(-k_{p}^2+\frac{q^2}{4}+k^2)}{qk} d z H^{(4)}_{np}(k,z,q)
+ \int_{k_{p}+\frac{q}{2}}^{k_{n}+\frac{q}{2}} k^2 dk \int_\frac{(k_{n}^2-\frac{q^2}{4}-k^2)}{qk}^{1} d z H^{(4)}_{np}(k,z,q)  \\
& + \int_{k_{n}+\frac{q}{2}}^{\infty} k^2 dk \int_{-1}^1 d z H^{(4)}_{np}(k,z,q) \Bigg],
\end{aligned}
\end{equation}
\begin{equation}
\begin{aligned}
H_{np}^{(1)}(k,z,q)=&\int d\boldsymbol{K} \mathcal{F}_{np}(k,z,K,Z,q)	\\
=& 2\pi \Bigg[\int_{k_{n}-\frac{q}{2}}^{k_{n}+\frac{q}{2}}  K^2 dK \int_\frac{(k_{n}^2-\frac{q^2}{4}-K^2)}{qK}^1 d Z \mathcal{F}_{np}(k,z,K,Z,q)  + \int_{k_{n}+\frac{q}{2}}^\infty   K^2 dK \int_{-1}^1 d Z \mathcal{F}_{np}(k,z,K,Z,q) \Bigg],
\end{aligned}
\end{equation}
\begin{equation}
\begin{aligned}
H_{np}^{(2)}(k,z,q)=&\int d\boldsymbol{K} \mathcal{F}_{np}(k,z,K,Z,q)	\\
=& 2\pi \Bigg[ \int_{\frac{1}{2}\sqrt{2k_{n}^2+2k_{p}^2-q^2}}^{k_{p}+\frac{q}{2}}  K^2 dK \int_\frac{(k_{n}^2-\frac{q^2}{4}-K^2)}{qK}^\frac{(-k_{p}^2+\frac{q^2}{4}+K^2)}{qK} d Z \mathcal{F}_{np}(k,z,K,Z,q) \\
&+ \int_{k_{p}+\frac{q}{2}}^{k_{n}+\frac{q}{2}} K^2 dK \int_\frac{(k_{n}^2-\frac{q^2}{4}-K^2)}{qK}^1 d Z \mathcal{F}_{np}(k,z,K,Z,q)
+ \int_{k_{n}+\frac{q}{2}}^\infty K^2 dK \int_{-1}^1 d Z \mathcal{F}_{np}(k,z,K,Z,q) \Bigg],
\end{aligned}
\end{equation}
\begin{equation}
\begin{aligned}
H_{np}^{(3)}(k,z,q)=&\int d\boldsymbol{K} \mathcal{F}_{np}(k,z,K,Z,q)	\\
=& 2\pi \Bigg[ \int_{k_{n}-\frac{q}{2}}^{-k_{p}+\frac{q}{2}}  K^2 dK \int_\frac{(k_{n}^2-\frac{q^2}{4}-K^2)}{qK}^1 d Z \mathcal{F}_{np}(k,z,K,Z,q)
+ \int_{-k_{p}+\frac{q}{2}}^{k_{p}+\frac{q}{2}} K^2 dK \int_\frac{(k_{n}^2-\frac{q^2}{4}-K^2)}{qK}^\frac{(-k_{p}^2+\frac{q^2}{4}+K^2)}{qK}  d Z \mathcal{F}_{np}(k,z,K,Z,q)  \\
 & + \int_{k_{p}+\frac{q}{2}}^{k_{n}+\frac{q}{2}} K^2 dK \int_\frac{(k_{n}^2-\frac{q^2}{4}-K^2)}{qK}^{1} d Z \mathcal{F}_{np}(k,z,K,Z,q)
+ \int_{k_{n}+\frac{q}{2}}^{\infty} K^2 dK \int_{-1}^1 d Z \mathcal{F}_{np}(k,z,K,Z,q) \Bigg],
\end{aligned}
\end{equation}
\begin{equation}
\begin{aligned}
H_{np}^{(4)}(k,z,q)=&\int d\boldsymbol{K} \mathcal{F}_{np}(k,z,K,Z,q)	\\
=& 2\pi \Bigg[ \int_{0}^{-k_{n}+\frac{q}{2}}  K^2 dK \int_{-1}^1 d Z \mathcal{F}_{np}(k,z,K,Z,q)
+ \int_{-k_{n}+\frac{q}{2}}^{-k_{p}+\frac{q}{2}}  K^2 dK \int_\frac{(k_{n}^2-\frac{q^2}{4}-K^2)}{qK}^1 d Z \mathcal{F}_{np}(k,z,K,Z,q)  \\
& + \int_{-k_{p}+\frac{q}{2}}^{k_{p}+\frac{q}{2}} K^2 dK \int_\frac{(k_{n}^2-\frac{q^2}{4}-K^2)}{qK}^\frac{(-k_{p}^2+\frac{q^2}{4}+K^2)}{qK} d Z \mathcal{F}_{np}(k,z,K,Z,q)
+ \int_{k_{p}+\frac{q}{2}}^{k_{n}+\frac{q}{2}} K^2 dK \int_\frac{(k_{n}^2-\frac{q^2}{4}-K^2)}{qK}^{1} d Z \mathcal{F}_{np}(k,z,K,Z,q)  \\
& + \int_{k_{n}+\frac{q}{2}}^{\infty} K^2 dK \int_{-1}^1 d Z \mathcal{F}_{np}(k,z,K,Z,q) \Bigg].
\end{aligned}
\end{equation}
Similarly, the integrands for the normalization and kinetic energy matrix elements are $\mathcal{F}_{np}(k,z,K,Z,q)=\hat{P}_{12}^{(np)}\phi^{*}_i(k,z,K,Z,q)\hat{P}_{12}^{(np)}\phi_j(k,z,K,Z,q)$, and $\hat{P}_{12}^{(np)}\phi_i^*(k,z,K,Z,q)\hat{T}_{4N}\hat{P}_{12}^{(np)}\phi_j(k,z,K,Z,q)$, respectively.
A 7-fold integral of integrand $\mathcal{F}_{np}(k,z,k',z',K,Z,q)=\hat{P}_{12}^{(np)}\phi_i^*(k,z,K,Z,q)\hat{V}\hat{P}_{12}^{(np)}\phi_j(k',z',K,Z,q)$ is required for the potential energy matrix element.

\end{widetext}

\end{document}